\documentstyle[12pt,amssymb]{article}
\begin{document}
\tolerance=5000
\def\pp{{\, \mid \hskip -1.5mm =}}
\def\cL{{\cal L}}
\def\be{\begin{equation}}
\def\ee{\end{equation}}
\def\bea{\begin{eqnarray}}
\def\eea{\end{eqnarray}}
\def\tr{{\rm tr}\, }
\def\nn{\nonumber \\}
\def\e{{\rm e}}

\newcommand{\vev}[1]{\left\langle#1\right\rangle}
\newcommand{\deri}[2]{\frac{d#1}{d#2}}
\newcommand{\pderi}[2]{\frac{\partial #1}{\partial #2}}
\newcommand{\ket}[1]{\left|#1\right\rangle}

\def\dfrac{\frac}

\begin{titlepage}

\ \hfill OCHA-PP-272

\begin{center}
\Large
{\bf Can the Baryon Number Density and\\the Cosmological Constant be interrelated?}

\bigskip
\bigskip

\normalsize

Azusa \textsc{Minamizaki}\footnote{E-mail address: minamizaki@hep.phys.ocha.ac.jp} 
and Akio \textsc{Sugamoto}\footnote{E-mail address: sugamoto@phys.ocha.ac.jp} 

\normalsize

\bigskip

{\em  Department of Physics, Ochanomizu University\\
Otsuka, Bunkyo-ku, Tokyo, Japan, 112-8610}

\end{center}

\vfill 

\baselineskip=24pt

\begin{abstract}
A toy model is proposed in which the cosmological constant and the baryon number density of the universe are interrelated. 
The model combines the mechanism of Dimopoulos and Susskind \cite{D-S} in which 
the baryon number density of the universe is generated by the time-dependence of the phase of a complex scalar field, \textit{i.e.} its `angular momentum' in the two-dimensional complex field space, 
with that of Yoshimura \cite{Yoshimura} in which the `centrifugal force' due to the `angular momentum' pushes the vacuum expectation value of the scalar
field out of a negative potential minimum and provides a small but positive cosmological constant.  Unfortunately, our model fails to relate the smallness of the two numbers directly, requiring a fine-tuning of the negative potential 
minimum.
\end{abstract}

\vfill

\end{titlepage}

\section{Introduction}

In this paper, we present a toy model of the universe
in which the baryon number density, $n_B$, and the cosmological 
constant, $\Lambda_{\mathrm{obs}}$, share a common dynamical origin.
As a consequence of originating from the same dynamics, $n_B$ and $\Lambda_{\mathrm{obs}}$ are interrelated in our model through the expression
\begin{equation}
\Lambda_{\mathrm{obs}} \;=\;-V_0 + \Lambda_1(n_B)\;,
\label{LambdaV0Lambda1}
\end{equation}
with
\begin{equation} 
\Lambda_1(n_B)=\left(\frac{a(T_B)}{a(T_0)}\right)^{-6}\cdot \left\{ \frac{1}{8q^2\vev{\phi_r^2}}\left({n_B}\right)^2 \right\}\;.
\label{LambdanBrelation}
\end{equation}
Here, $a(T)$ is the scale of the universe at temperature $T$,
with $T_0 = 2.7\,{\rm K} = 2.3\times 10^{-4}\,{\rm eV}$ 
being the current temperature,
and $T_B\approx 1\,{\rm TeV}$ being the temperature at which electroweak baryogenesis is assumed to have taken place;
$q$ is a parameter which we equate to the number of fermion generations $N_g$, $\vev{\phi_r}$ is the vacuum expectation value (VEV) 
of a scalar field in our model, and $-V_0$ is the depth of the potential well
that the scalar field is subjected to.

Ideally, one would want an expression such as Eq.~(\ref{LambdanBrelation})
to not merely interrelate $n_B$ and $\Lambda_{\mathrm{obs}}$, but to also
provide a natural explanation of why $\Lambda_{\mathrm{obs}}$ is so small
by relating it directly to the smallness of $n_B$.  Indeed, the values of
$n_B$ and $\Lambda_{\mathrm{obs}}$ as determined by WMAP \cite{WMAP} are:
\begin{eqnarray}
\Lambda_{\mathrm{obs}} 
&\approx& 30 \times ({\rm meV})^4 
\;=\; 30 \times 10^{-48}({\rm GeV})^4\;, \cr
n_{B}
&\approx& 2.5\times 10^{-7}/{\rm cm}^3
\;\approx\; 2 \times 10^{-48}({\rm GeV})^3\;,
\end{eqnarray}
sharing the suggestive factor of $10^{-48}$ when expressed in GeV units,
hinting that such a relation may exist.
Unfortunately, Eq.~(\ref{LambdanBrelation}) does not have this property:
if we take $\vev{\phi_r}=O(10^2\,\mathrm{GeV})$, 
we find $\Lambda_1 = O(10^{-8}\,\mathrm{GeV}^4)$ for the above value of
$n_B$, which necessitates the extreme fine-tuning of the $-V_0$ term 
in Eq.~(\ref{LambdaV0Lambda1}) to obtain 
the desired value for $\Lambda_{\mathrm{obs}}$.

Therefore, our toy model does not solve the cosmological constant problem. 
Nevertheless, the mechanism that leads to 
Eq.~(\ref{LambdanBrelation}) is new, and
the main objective of this paper is its presentation. 
This paper is organized as follows:
In section 2, we briefly review
the models proposed by Dimopoulos and Susskind \cite{D-S}, 
and by Yoshimura \cite{Yoshimura}, and outline how their mechanisms 
are combined into one in our toy model.
In section 3, we introduce our model and derive the relation between
the baryon number density and the cosmological constant.
Section 4 concludes with a discussion on how one may 
improve upon our model so that the smallness of
$n_B$ and $\Lambda_{\mathrm{obs}}$ can be directly related.

\section{Dimopoulos-Susskind and Yoshimura Models}

Since our toy model combines the mechanism found in the model 
proposed by Dimopoulos and Susskind \cite{D-S}, 
which generates a non-zero baryon number density dynamically, 
with that found in the model proposed by Yoshimura \cite{Yoshimura},
which generates a non-zero cosmological constant dynamically,
we begin by giving a brief description of these mechanisms.

In the Dimopoulos-Susskind model \cite{D-S}, 
a complex scalar field, $\phi = |\phi| e^{i\theta}$, 
which carries baryon number, is subjected to 
a B, C, and CP violating potential:
\begin{equation}
V_{\mathrm{DS}}(\phi) 
= \dfrac{\lambda}{M^{2n}}\,(\phi^*\phi)^n (\phi+\phi^*)(\alpha\phi^3+\alpha^*\phi^{*3})\;,
\end{equation}
where $\lambda$ is a dimensionless coupling constant,
$M$ is a mass scale, $n$ is some integer, and $\alpha$ is a constant phase.
This potential is periodic but left-right asymmetric in the 
$\theta$-direction for fixed $|\phi|$.
This asymmetry, along with loop effects which provide
`friction' to the time-evolution of the field, leads to the phase of the
complex field being driven preferentially in one direction.\footnote{%
Without friction, $\dot{\vev{\theta}}$ will average out to zero.} 
Since the baryon number current is proportional to $\partial_\mu\theta$,
this results in baryon number violation:
\begin{equation}
n_B \propto \dot{\vev{\theta}} \neq 0\;.
\end{equation}
In other words, in the Dimopoulos-Susskind model, the baryon number density is
generated by the `angular velocity' of the complex scalar field 
in the 2-dimensional complex field space.
Of course, this type of `baryon number' density is not exactly what we mean
by the term.  To remedy this problem, we can couple the scalar field
to a fermionic baryon number current by a term of the form
\begin{equation}
\left(\partial_\mu\theta\right) \left(\bar{\psi}\gamma^\mu\psi\right)\;,
\end{equation}
as in the spontaneous baryogenesis mechanism by Cohen and Kaplan \cite{C-K},
which has the effect of converting the bosonic baryon number into a fermionic baryon number.  We will have more to say about this later.

Yoshimura \cite{Yoshimura} proposed a Brans-Dicke type gravitational model with 
two dilaton fields which generates a small cosmological constant dynamically.
The two real dilaton fields can be combined into a single complex scalar field 
$\phi = |\phi|e^{i\theta}$, and it is subjected to the potential
\begin{equation}
V_{\mathrm{Yoshimura}}(\phi) = V_0\,\cos\dfrac{|\phi|}{M}\;,
\end{equation}
where $M$ is a mass scale assumed to of the order of a TeV, and
$V_0$ is a constant of order $M^4$.
In the Yoshimura model, as the universe cools, $|\phi|$ settles down
into one of the negative potential minima of $-V_0$, 
but $\dot\theta$ stays non-zero, the `kinetic energy' of this 
`angular motion' contributing positively to the cosmological constant.
This kinetic energy contribution 
is assumed to be of order $M^4$, allowing for a cancellation
against the potential term $-V_0$.  It should be noted, though, that
extreme fine-tuning is still required for this cancellation to be exact.

In our toy model, we combine these two mechanisms: the 
`baryon number' generated {\it a la} Dimopoulos and Susskind provides the `angular motion' of Yoshimura, and a negative potential minimum is enhanced by this `angular motion' to a very small value of the observed cosmological constant {\it a la} Yoshimura.  In this way, the baryon number and the cosmological constant are interrelated in our toy model.

To understand the relationship between the baryon number density and the cosmological constant in our model, it is helpful to visualize in our minds
the dynamics of a point particle in a two-dimensional central force potential 
$V(r)$.  
If we consider the particle following a circular orbit, the radius of the
orbit and the energy of the particle are determined by the minimum of 
the effective potential:
\begin{equation}
U_{\mathrm{eff}}(r) = V(r) + \dfrac{L^2}{2mr^2}\;,
\end{equation}
where $m$ is the mass, and $L$ is the angular momentum of the particle.
If the depth and shape of the potential $V(r)$ are such that it
cancels the centrifugal barrier term $L^2/(2mr^2)$, 
the minimum of $U_{\mathrm{eff}}(r)$ could be very small and positive.
The mechanism in which the cosmological constant is generated in our toy model is just a field theoretical version of this well-known mechanism, being performed in the expanding universe, where the baryon number density is a 
field-theoretical analogue of the angular momentum. \\

\section{The Toy Model} 

We assume the existence of a fermionic sector whose Lagrangian is 
invariant under the following global phase transformation with respect to a parameter $\alpha$: 
\begin{equation}
\psi_i \;\longrightarrow\; e^{+ib_i\alpha}\,\psi_i\;,\qquad 
\bar{\psi}_i \;\longrightarrow\; e^{-ib_i\alpha}\,\bar{\psi}_i\;. 
\label{fermionic transformation}
\end{equation}
Here, $b_i$ is the baryon number, B, of fermion $\psi_i$.
This invariance implies the conservation of the baryon number current given by
\begin{equation}
j_B^\mu\;=\;\sum_{i}b_i\left(\bar{\psi}_i\gamma^{\mu}\psi_i\right)\;. 
\end{equation}
Next, we introduce a complex scalar field 
$\phi(x)=\phi_{r}(x)e^{i\theta(x)}$ with baryon number $q$,
\textit{i.e.} it transforms as 
\begin{equation}
\phi\;\longrightarrow\; e^{+iq\alpha}\,\phi\;,\qquad  
\phi^*\;\longrightarrow\; e^{-iq\alpha}\,\phi^*\;,
\label{scalar transformation}
\end{equation}
under the B-transformations. The associated current is
\begin{equation}
j_\phi^\mu 
\;=\; -q\Bigl(\phi^* \,i\!\stackrel{\leftrightarrow}{\partial^\mu}\!\phi\Bigr) 
\;=\; 2\,q\,\phi_r^2\,\partial^\mu \theta\;.
\end{equation}
Note that the time component of this current is given by the angular velocity of the phase of the complex scalar field.  This is the `baryon number' current of the Dimopoulos-Susskind model. 
In addition to the fermionic and bosonic sectors, we introduce an interaction between the fermions (baryons) and the boson (complex scalar) via the Cohen-Kaplan term \cite{C-K}, 
\begin{equation}
\dfrac{1}{q}\left(\partial_{\mu}\theta\right) j_{B}^{\mu}\;,
\label{CKterm}
\end{equation}
where $q$ is the B-charge of the complex scalar.
In the absence of B-violating potentials, the combined
fermionic and bosonic sectors, together with the interaction term, 
is invariant under the combined transformations 
Eq.~(\ref{fermionic transformation}) and Eq.~(\ref{scalar transformation}), 
so that the sum of the fermionic and bosonic currents is conserved:
\begin{equation}
\partial_\mu \left(\, j_B^\mu + j_{\phi}^\mu \,\right)\;=\;0\;.
\end{equation}

At this point, we would like to discuss the meaning of the
interaction term, Eq.~(\ref{CKterm}).
If the reader is familiar with PCAC \cite{PCAC}, the term can be understood as 
follows: If the symmetry of baryon number conservation (like chiral symmetry) is spontaneously broken, the Nambu-Goldstone (NG) mode $\theta$ 
(like the $\pi$ meson) appears so that the conservation law may be restored. 
The NG mode couples derivatively to the fermions via the Cohen-Kaplan term, 
just like the pion-nucleon coupling $g_{NN\pi}({\bar N}\gamma^{\mu}\gamma_5 N )\partial_{\mu}\pi$.  
It is in this sense that Cohen and Kaplan called the mechanism of their model `spontaneous baryogenesis'.

Another way to understand the term may be possible from the viewpoint of the sphaleron \cite{sphaleron}: 
The symmetry of baryon number conservation is broken by the anomaly, but the topological Chern-Simons current appears from the degrees of freedom of gauge fields, which restores the conservation.  Then, the sum of the fermionic current and the bosonic Chern-Simons (CS) current, $k_{\mathrm{CS}}^{\mu}$, is conserved
\begin{equation}
\partial_\mu \left(\, j_B^\mu + N_g \cdot k_{\mathrm{CS}}^\mu \,\right)\;=\;0\;,
\end{equation}
where the interaction is induced by the anomaly.
If the scalar current in our toy model could be considered as a simplified version of this CS current, then $q$ becomes the number of generations $N_g$, since in the sphaleron model, one unit of Chern-Simmons number corresponds to $N_g$ baryon numbers.

In our toy model, the conserved quantity (the `baryon number density') is the sum of the fermionic baryon number density $n_B$ and the `angular velocity' of the complex scalar phase.  
Therefore, even if we start from zero total `baryon number', 
we can generate the fermionic baryon number density $n_B$ from the 
`angular velocity' of the complex scalar phase as follows:
\begin{equation}
n_B\;=\;-2\,q\,\phi_r^2\,\dot{\theta}\;.
\end{equation}

We place our model in an expanding universe which can be described by the Robertson-Walker metric
\begin{equation}
\left(ds\right)^{2}\;=\;\left(dt\right)^{2}-\left(\frac{a\left(t\right)}{a_0}\right)^{2}\left(d\vec{x}\right)^{2},
\end{equation}
where $a_0=a(t_0)$ is the scale factor at the present time.
The action of our toy model is then
\begin{equation}
S\;=\;\int d^{4}x \sqrt{-g}
\left[\,
g^{\mu\nu} \partial_{\mu} \phi^{*}\partial_{\nu}\phi
-V\left(\phi,\phi^*\right)
+\dfrac{1}{q}\,g^{\mu\nu} (\partial_{\mu}\theta) j_{B\nu}
+{\cal L}_m 
\,\right]\;,
\end{equation}
where ${\cal L}_m$ is the Lagrangian for the matter sector (fermions and other particles in the Standard Model).  
If the action is rewritten in terms of $\phi_r$ and $\theta$, and if the spacial derivatives are ignored under the assumption of spacial uniformity, the action becomes
\begin{equation}
S=\int d^{4}x \left(\frac{a\left(t\right)}{a_0}\right)^{3}
\left[\,(\dot{\phi}_{r})^{2}
+\phi_{r}^{2}\;\dot{\theta}^{2}
-V\left(\phi_{r},\theta \right) 
+\dfrac{1}{q}\,\dot{\theta}\,n_{B}
+{\cal L}_m \,\right].
\end{equation}
We separate the scalar potential, $V(\phi_{r}, \theta)$ into a B, C and CP conserving part $V_{r}(\phi_{r})$, and a B, C and CP violating part 
$V_{\theta}(\phi_{r}, \theta)$:
\begin{equation}
V(\phi_{r}, \theta)=V_{r}(\phi_{r})+V_{\theta}(\phi_{r}, \theta).
\end{equation} 
There are many possibilities for the B, C and CP conserving potential
$V_r(\phi_r)$, its detailed shape being irrelevant for our discussion.
The simplest one is the usual wine-bottle potential,
\begin{equation}
V_{\rm wine-bottle}(\phi)= -\mu^2 (\phi^{*}\phi) + \lambda(\phi^{*}\phi)^2= -V_0 + \lambda \left(\phi_{r}-\langle\phi_{r}\rangle \right)^4.
\end{equation}
It is also possible to choose $V_r(\phi_r)$ to be the one used 
to understand inflation, $V_{\rm inflation}(\phi)$, which takes a positive value near $\phi_{r}=0$, but modify it to take on a negative minimum value $-V_0$ at $\phi_{r}=\langle\phi_{r}\rangle$.
The third possibility is the periodic potential proposed by Yoshimura,
\begin{equation}
V_{\mathrm{Yoshimura}}(\phi)=V_{0}\cos \frac{\phi_{r}}{M}\;,
\end{equation}
where $M$ gives the period of the potential and the minimum value is $-V_0$.
We assume that after a long passage of time, $\phi_{r}$ loses its energy and relaxes and settles into a minimum of $V_r(\phi_r)$ at 
$\langle\phi_{r}\rangle$, where the potential minimum gives the negative value $-V_0$.

An example of a B, C and CP violating potential is given by Dimopoulos and Susskind \cite{D-S}, which we adopt, namely,
\begin{eqnarray}
V_{\theta}(\phi)&=&\lambda' \left(\phi+\phi^{*}\right)\left(\alpha \phi^{3}+\alpha^{*}\phi^{*3}\right),\\
&=&\lambda' \left( \phi_{r}\right)^4 4\cos\theta\cdot\cos\left(3\theta+\beta\right),
\end{eqnarray}
where $\alpha=e^{i\beta}$.  This potential manifestly violates B (the symmetry of rotation, or the shift of $\theta$) as well as C and CP, 
since the C-transformation for $\phi$ is $\phi (t, x)\to \phi^*(t, x)$ 
and the CP-transformation is $\phi  (t, x)\to \phi^* (t, -x)$.
This B, C and CP violating potential is periodic and has many maxima and minima in the $\theta$-direction.  These maxima and minima may
come from fluctuations of the scalar field, or from the maxima and minima apparent in the sphaleron transition.

It is reasonable to assume that the CP-violating quartic coupling $\lambda'$ is very small compared to CP-conserving couplings such as $\lambda$, so that even after $\phi_{r}$ relaxes to the vicinity of a minimum of the CP-conserving potential $V_r$, the system has enough energy to let $\theta$ move smoothly over the small bumps of the potential $V_{\theta}$.  
Therefore, except for the small bumps, the potential is approximately rotationally symmetric in the two dimensional complex $\phi$ space, and the analogue of `angular momentum' is approximately conserved, 
which provides a `centrifugal force' which pushes $\vev{\phi_r}$ 
away from its potential minimum.  
The final vacuum energy is determined from the competition of this `centrifugal force' and the `centripetal force' from the radial potential $V_r$. 

The potential in the $\theta$-direction is identical to the Dimopoulos-Susskind model, so we can expected it to lead to a non-zero angular velocity:  
Because the potential is CP-violating, \textit{i.e.} asymmetric with respect to the reflection $\theta \to -\theta$, accelerations in the forward and backward directions are asymmetric.  
Furthermore, the decay width of the scalar to fermions ($\dot\theta\to$ fermion pairs) through the Cohen-Kaplan term plays the role of a damping term in the equation of motion for $\dot{\theta}$.  
It should be noted, however, that the generation of a non-zero 
`time-averaged angular momentum' is not guaranteed by the presence of 
CP-asymmetry and friction.
In fact, in the original Dimopoulos-Susskind model without fermions, 
non-renormalizable interactions are required in order to obtain a nonvanishing angular velocity: the potential needs to be modified to $V_{\theta} \rightarrow (\phi_{r}/M)^{n} \times V_{\theta}$, where $n$ is a positive integer.  
However, it may be possible to generate a 
non-zero velocity for $\theta$ using other ideas such as those
used in the theory of molecular motors, or ratchets, where the generation of a steady rotation is also necessary.  
For example, a non-zero $\dot{\theta}$ may be obtained even with the renormalizable interaction adopted in our model, if the amplitude of the potential $V_{\theta}$ is changed periodically in time (like in the parametric resonance of a pendulum).  
Here, the amplitude is $\lambda (\phi_{r})^4$, so the thermal oscillation of 
$\phi_{r}$ around the minimum position of $V_{r}$ at the final stage of inflation may be utilized.  
Detailed analysis on whether the desired `angular velocity' can be generated this way is left to a future study \cite{ratchet}.

Here, we will only survey the implications when the desired angular velocity is obtained in our toy model, and derive a formula relating the two small numbers, the angular momentum (the baryon number density $n_B$) and the vacuum energy (the observed cosmological constant $\Lambda_{obs}$). 
The equation of motion for $\theta$ in our model is 
\begin{equation}
\ddot{\theta}+3H\dot{\theta}
+2\,\dfrac{\dot{\phi_r}}{\phi_r}\,\dot{\theta}
+\dfrac{1}{2\phi_r^2}\,\dfrac{\partial V}{\partial\theta}
\;=\;-\dfrac{1}{2q\phi_r^2}\left(\dot{n}_B+3Hn_B\right)\;.
\end{equation}
After enough time has elapsed, 
$\ddot{\theta}$, $\dot{\phi}_r$, and $\dot{n}_B$ all go to zero, but the
average values of $\dot{\theta}$, $\phi_r$, and $n_B$ are assumed to relax and settle to certain constants.   
Averaging the above equation of motion over periods of $\theta$ around the baryon number generation time $t_B$, or the temperature $T_B$, we obtain
\begin{equation}
\vev{\dot{\theta}(t_B)}=-\frac{1}{2q\vev{\phi_r^2}}\vev{n_B(t_B)}.
\end{equation}
This gives a relation between the `angular velocity' and the baryon number density at $t_B$.  Here, we have specified the time to be $t_B$, 
or equivalently, the temperature to be $T_B$, at which baryon number is generated, since it is only around this time or temperature that the angular velocity exists, and baryon number can be generated.

The energy density of the universe, $\varepsilon$, includes the kinetic energy of the `rotational motion' and reads
\begin{equation}
\varepsilon(t)\;=\;
{\mathcal{H}}\;=\;
\left(\dfrac{a(t)}{a_0}\right)^3
\left\{ \dfrac{1}{2} \,\phi_r^2\,\dot{\theta}(t)^2
+ \dfrac{1}{2}\dot{\phi}_r^2
+ V(\phi_r,\theta)\right\}
\;,
\end{equation}
where we have absorbed the zero-point fluctuation energies of the SM matter fields into the potential $V(\phi_r,\theta)$.
Again, taking the time average over rotation periods around the baryogenesis time $t_B$, we find the following relation between the baryon number density and the energy density of the universe both at $t_B$:
\begin{equation}
\varepsilon(t_B)={\left(\frac{a(t_B)}{a_0}\right)}^3\left\{-V_0 + \frac{1}{8q^2\vev{\phi_r^2}}{\vev{n_B(t_B)}}^2\right\}\;.  
\label{A}
\end{equation}
Here, $-V_0$ is the value of the potential at $\vev{\phi_r}$.
The baryon number generated at $t_B$, or $T_B$, is diluted with 
the expansion of the universe.  That is, the baryon number density at the present time, $n_B$, is given by
\begin{equation}
n_B={\left(\frac{a(t_B)}{a_0}\right)}^3 n_B(t_B)\;. 
\label{B}
\end{equation}
On the other hand, the energy density of the universe at $t_B$ is related to the cosmological constant $\Lambda_{\rm obs}$ as
\begin{equation}
\varepsilon(t_B)={\left(\frac{a(t_B)}{a_0}\right)}^3 \cdot \Lambda_{\rm obs}\;,  \label{C}
\end{equation}
since the cosmological constant is defined by the following action 
\begin{equation}
S= \int d^4 x \sqrt{-g} \cdot \Lambda_{\rm obs}\;. 
\end{equation}
From Eqs.~(\ref{A})-(\ref{C}), we obtain the following simple relation between the current baryon number density $n_B$, and the observed cosmological constant $\Lambda_{\rm obs}$, 
\begin{equation}
\Lambda_{\rm obs}=-V_0 + \left\{ \left( \frac{a(t_B)}{a_0} \right) \right\} ^{-6} \cdot 
\left\{ \frac{1}{8q^2\vev{\phi_r^2}}\left(n_B \right)^2 \right \}\;,
\end{equation}
where ${a(t_B)}/{a_0}$ can also be expressed as ${T_0}/{T_B}$.

This is the formula which connects the two small numbers, 
the baryon number density and the cosmological constant.  
As was discussed in the introduction, if we substitute the observed values of $n_B \approx 2 \times 10^{-48}({\rm GeV})^3$ and $\Lambda_{\rm obs} \approx 30\times 10^{-48}({\rm GeV})^4$,
as well as the values of $T_B=1\,{\rm TeV}$, 
$\vev{\phi_{r}}=174\,{\rm GeV}$, and 
$q=N_g$ (number of generations)$=3$, we obtain 
\begin{equation}
\Lambda_{obs}\;\approx\; 
30\times 10^{-48}({\rm GeV})^4
\;=\;-V_0+1.2 \times 10^{-8} ({\rm GeV})^4\;.
\end{equation}
Here, $T_B$ and $\vev{\phi_r}$ are, respectively, the energy scale at which baryon number is generated, and the energy scale of the model.  
As is evident, 
in order to obtain a small cosmological constant within our model, 
fine-tuning of $V_0$ is necessary.  Nevertheless, 
our model does successfully relate the cosmological constant to the
baryon number density via a very simple formula.

\section{Discussion}

We would like to begin this discussion by presenting what originally
led us to construct our toy model.

The observation of WMAP \cite{WMAP} is consistent with 
the $\Lambda$CDM model of the Universe, namely that with Cold Dark Matter, and a non-zero cosmological constant $\Lambda$.  The fit to the data gives 
\begin{equation}
\Lambda_{\mathrm{obs}} 
\;\approx\; 30 \times ({\rm meV})^4 
\;=\; 30 \times 10^{-48}({\rm GeV})^4\;.  
\label{LambdaValue}
\end{equation}
From the same observation, the baryon-to-photon ratio
$\eta \equiv n_{B}/n_{\gamma}$ is determined to be
\begin{equation}
\eta \;=\; (6.14\pm 0.25)\times 10^{-10}\;.
\end{equation}
As the photon number density $n_{\gamma}$ is about $410/{\rm cm}^3$, the baryon number density becomes  
\begin{equation}
n_{B}
\;\approx\; 2.5\times 10^{-7}/{\rm cm}^3
\;\approx\; 2 \times 10^{-48}({\rm GeV})^3\;.
\label{nBValue}
\end{equation}
As mentioned in the introduction, 
both the cosmological constant and the baryon number density have 
a suggestive common factor of $10^{-48}$ when expressed in units of GeV 
to their respective powers. 
The same factor appears in the ratio of the current three-dimensional volume of the Universe (at temperature $T_0 = 2.7\,{\rm K} = 2.3\times 10^{-4}\,{\rm eV}$), to what it was at the electroweak baryogenesis temprature 
$T_B\approx 1\,{\rm TeV}$:
\begin{equation}
\left(\dfrac{a(T_B)}{a(T_0)}\right)^3
\;\approx\; \left(\dfrac{2.3\times 10^{-4}\,\mathrm{eV}}{10^{12}\,\mathrm{eV}}\right)^3
\;=\; 12\times 10^{-48}\;.
\label{aRatioValue}
\end{equation}
Here, we have used $a(t) = a(T) \propto 1/T$, \textit{i.e.} replaced the
time-dependence of the scale of the Universe with its 
temperature-dependence \cite{coincide}.

These coincidences are reminiscent of those that led Dirac to postulate
his famous Large Number Hypothesis \cite{Dirac}, \textit{i.e.}, very large dimensionless numbers (or very small ones if you take the reciprocal) that appear in nature are somehow all interrelated. In his case, the common factor was $10^{39}$, which was the age of the Universe measured in units of $e^2/m_e c^3$.
Following the spirit of Dirac, we postulated that the cosmological constant
and the baryon density of the Universe are interrelated through the
baryogenesis mechanism.
In contrast to the Dirac case, $\Lambda$ and $n_B$ are
dimension-\textit{ful} quantities, and the coincidence in their numerical values 
occur only if expressed in GeV units.
This scale must be set by electroweak baryogenesis which occurs at 
$\sim 1\,\mathrm{TeV}=10^3\,\mathrm{GeV}$.

So our original objective 
was to construct a toy model which relates the three numbers
of Eqs.~(\ref{LambdaValue}), (\ref{nBValue}), and (\ref{aRatioValue}) 
in such a way that
the common factor of $10^{-48}$ cancels in the relation, providing
an explanation of why $\Lambda_{\mathrm{obs}}$ is so small by relating it directly to the smallness of $n_B$.
Our current model clearly misses this mark.

Let us speculate on possible scenarios in which our original idea may work.
Suppose that we could find a model which predicts a formula of the form.
\begin{equation}
\Lambda_{\rm obs} 
\;\propto\; {\left(\frac{a(T_B)}{a(T_0)}\right)}^{-3x}\left\{\frac{\left({n_B}\right)^{(x+1)}}{\vev{\phi_r}^{(3x-1)}}\right\}\;,
\end{equation} 
where $x$ is a real number.  Then, the cosmological constant of order $10^{-48}({\rm  GeV})^4$ could be derived from the baryon number density of order $10^{-48}({\rm  GeV})^3$, if it is generated at $T_B$ in a theory with $\vev{\phi_{r}}$ of the order of GeV or TeV.

The case with $x=1$ would be realized in a theory with the fractal spacial dimension $D=3/2$, \textit{i.e.}
the baryon number is diluted by the expansion of the Universe 
as if it were contained in a $D=3/2$ dimensional can.  
The case with $x=2$ would be realized if the `angular momentum' $\mathbf{L}$ contributed to the energy as $\mathrm{L}^3$. 
This can be realized in a scalar model in which the angular momenta are in a triplet representation $\mathbf{L}^{a}$ $(a=1,2,3)$, and the Hamiltonian is proportional to $\det(L_i^a)$.  This case may not be impossible.  
Another possibility is that if the cosmological constant would change in time like $a(t)^3$, then the cosmological constant observed at present could be properly predicted without fine tuning.  
We intend to pursue these ideas in a future publication.  

\ 

\noindent{\bf Acknowledgments}

We give our sincere thanks to Tatsu Takeuchi for valuable discussions and for suggesting the connection between our model and molecular motors.  
A.M. would also like to thank him for helping her with her English when she gave a talk at the Joint Meeting of Pacific Particle Physics Communities in Hawaii, November 2006.  
A.S. is partially supported by a Grant-in-Aid for Scientific Research (No.17540238) from the Ministry of Education, Culture, Sports, Science and Technology of Japan.

\end{document}